# Incorporating biological structure into machine learning models in biomedicine

## Authors

- **Jake Crawford**
  [0000-0001-6207-0782](#) · [jjc2718](#) · [jjc2718](#)
  Graduate Group in Genomics and Computational Biology, Perelman School of Medicine, University of Pennsylvania, Philadelphia, PA; Department of Systems Pharmacology and Translational Therapeutics, Perelman School of Medicine, University of Pennsylvania, Philadelphia, PA · Funded by National Human Genome Research Institute (R01 HG010067)

- **Casey S. Greene**
  [0000-0001-8713-9213](#) · [cgreene](#) · [greenescientist](#)
  Department of Systems Pharmacology and Translational Therapeutics, Perelman School of Medicine, University of Pennsylvania, Philadelphia, PA; Childhood Cancer Data Lab, Alex's Lemonade Stand Foundation, Philadelphia, PA · Funded by The Gordon and Betty Moore Foundation (GBMF 4552), National Human Genome Research Institute (R01 HG010067), National Cancer Institute (R01 CA237170), Alex's Lemonade Stand Foundation (CCDL)

## Abstract

In biomedical applications of machine learning, relevant information often has a rich structure that is not easily encoded as real-valued predictors. Examples of such data include DNA or RNA sequences, gene sets or pathways, gene interaction or coexpression networks, ontologies, and phylogenetic trees. We highlight recent examples of machine learning models that use structure to constrain model architecture or incorporate structured data into model training. For machine learning in biomedicine, where sample size is limited and model interpretability is critical, incorporating prior knowledge in the form of structured data can be particularly useful. The area of research would benefit from performant open source implementations and independent benchmarking efforts.

## Graphical abstract

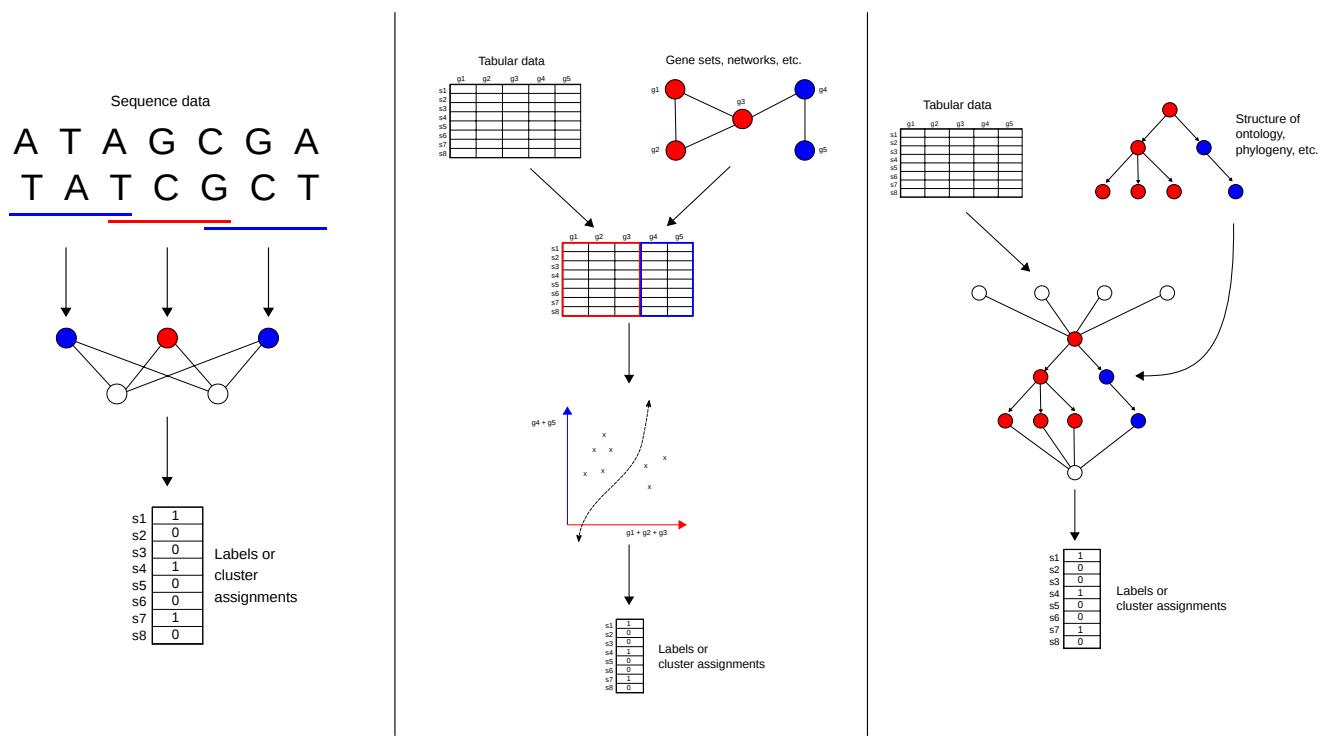

Schematic showing the main categories of models incorporating structured biological data covered in this review. The first panel shows an example of a model operating on raw sequence data, the second panel shows a model in which dimension reduction is constrained by the connections in a gene network, and the third panel shows a neural network with structure constrained by a phylogeny or ontology.

## Introduction

It can be challenging to distinguish signal from noise in biomedical datasets, and machine learning methods are particularly hampered when the amount of available training data is small. Incorporating biomedical knowledge into machine learning models can reveal patterns in noisy data [1] and aid model interpretation [2]. Biological knowledge can take many forms, including genomic sequences, pathway databases, gene interaction networks, and knowledge hierarchies such as the Gene Ontology [3]. However, there is often no canonical way to encode these structures as real-valued predictors. Modelers must creatively decide how to encode biological knowledge that they expect will be relevant to the task.

Biomedical datasets often contain more input predictors than data samples [4,5]. A genetic study may genotype millions of single nucleotide polymorphisms (SNPs) in thousands of individuals, or a gene expression study may profile the expression of thousands of genes in tens of samples. Thus, it can be useful to include prior information describing relationships between predictors to inform the representation learned by the model. This contrasts with non-biological applications of machine learning, where one might fit a model on millions of images [6] or tens of thousands of documents [7], making inclusion of prior information unnecessary.

We review approaches that incorporate external information about the structure of desirable solutions to learn from biomedical data. One class of commonly used approaches learns a representation that considers the context of each base pair from raw sequence data. For models that operate on gene expression data or genetic variants, it can be useful to incorporate networks or pathways describing relationships between genes. We also consider other examples, such as neural network architectures that are constrained based on biological knowledge.

There are many complementary ways to incorporate heterogeneous sources of biomedical data into the learning process, which have been covered elsewhere [8,9]. These include feature extraction or representation learning prior to modeling and/or other data integration methods that do not necessarily involve customizing the model itself.

## Sequence models

Early neural network models primarily used hand-engineered sequence features as input to a fully connected neural network [10,11] (Figure 1). As convolutional neural network (CNN) approaches matured for image processing and computer vision, researchers leveraged biological sequence proximity similarly. CNNs are a neural network variant that groups input data by spatial context to extract features for prediction.

The definition of "spatial context" is specific to the input: one might group image pixels that are nearby in 2D space, or genomic base pairs that are nearby in the linear genome. In this way, CNNs consider context without making strong assumptions about exactly how much context is needed or how it should be encoded; the data informs the encoding. A detailed description of how CNNs are applied to sequences can be found in Angermueller et al. [12].

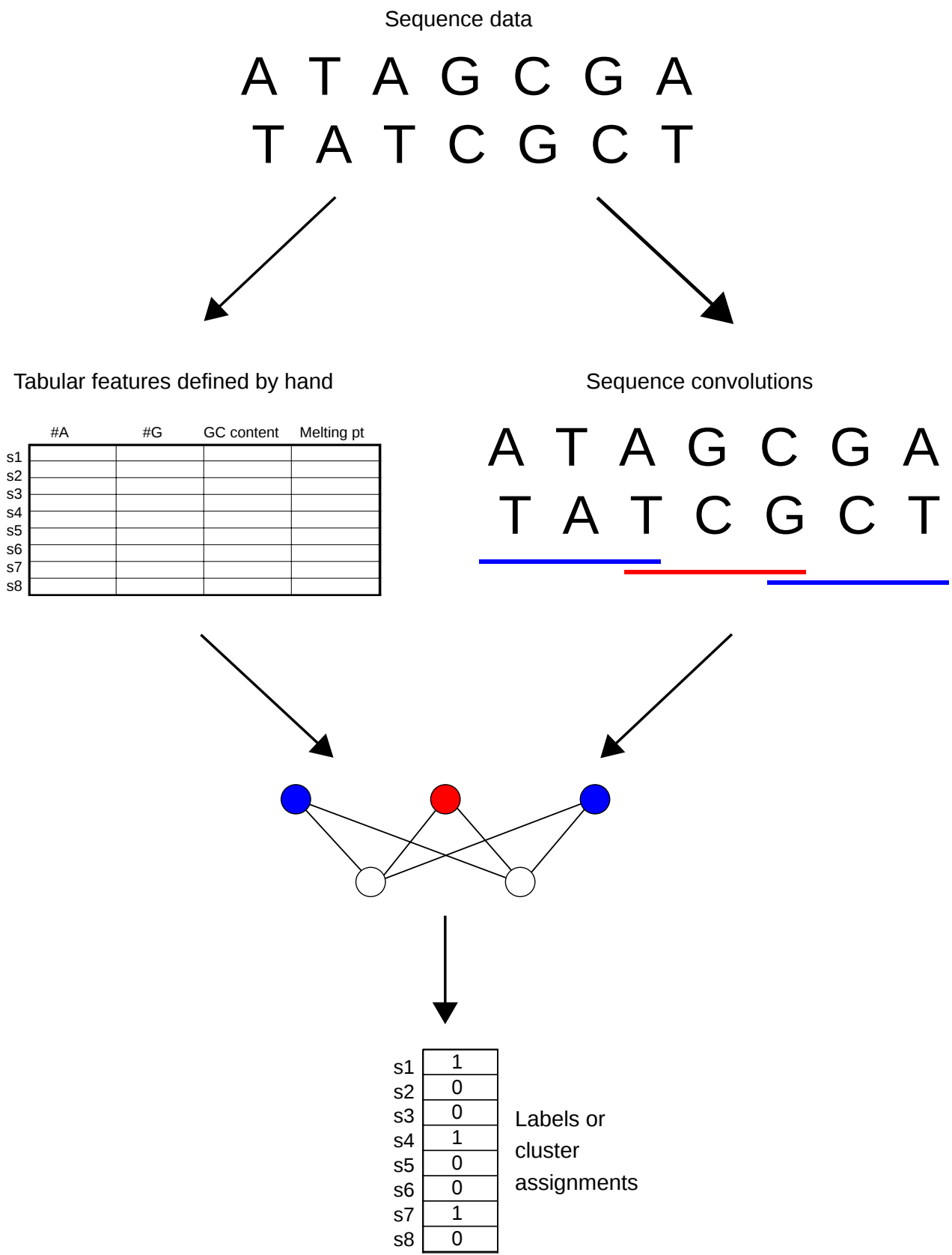

**Figure 1:** Contrasting approaches to extracting features from DNA or RNA sequence data. Early models defined features of interest by hand based on prior knowledge about the prediction or clustering problem of interest, such as GC content or sequence melting point. Convolutional models use sequence convolutions to derive features directly from sequence proximity, without requiring features of interest to be identified before the model is trained.

## Applications in regulatory biology

Many early applications of deep learning to biological sequences were in regulatory biology. Early CNNs for sequence data predicted binding protein sequence specificity from DNA or RNA sequence [13], variant effects from noncoding DNA sequence [14], and chromatin accessibility from DNA sequence [15].

Recent sequence models take advantage of hardware advances and methodological innovation to incorporate more sequence context and rely on fewer modeling assumptions. BPNet, a CNN that predicts transcription factor binding profiles from DNA sequences, accurately mapped known locations of binding motifs in mouse embryonic stem cells [16]. BPNet considers 1000 base pairs of context around each position when predicting binding probabilities with a technique called dilated convolutions [17], which is particularly important because motif spacing and periodicity can influence binding. cDeepbind [18] combines RNA sequences with information about secondary structure to predict RNA binding protein affinities. Its convolutional model acts on a feature vector combining sequence and structural information, using context for both to inform predictions. APARENT [19] is a CNN that predicts alternative polyadenylation (APA) from a training set of over 3 million synthetic APA reporter sequences. These diverse applications underscore the power of modern deep learning models to synthesize large sequence datasets.

Models that consider sequence context have also been applied to epigenetic data. DeepSignal [20] is a CNN that uses contextual electrical signals from Oxford Nanopore single-molecule sequencing data to predict 5mC or 6mA DNA methylation status. MRCNN [21] uses sequences of length 400, centered at CpG sites, to predict 5mC methylation status. Deep learning models have also been used to predict gene expression from histone modifications [22,23]. Here, a neural network model consisting of long short-term memory (LSTM) units was used to encode the long-distance interactions of histone marks in both the 3' and 5' genomic directions. In each of these cases, proximity in the linear genome helped model the complex interactions between DNA sequence and epigenome.

## Applications in variant calling and mutation detection

Identification of genetic variants also benefits from models that include sequence context. DeepVariant [24] applies a CNN to images of sequence read pileups, using read data around each candidate variant to accurately distinguish true variants from sequencing errors. CNNs have also been applied to single molecule (PacBio and Oxford Nanopore) sequencing data [25], using a different sequence encoding that results in better performance than DeepVariant on single molecule data. However, many variant calling models still use hand-engineered sequence features as input to a classifier, including current state-of-the-art approaches to insertion/deletion calling [26,27]. Detection of somatic mutations is a distinct but related challenge to detection of germline variants, and has also recently benefitted from use of CNNs [28].

## Network- and pathway-based models

Rather than operating on sequences, many machine learning models in biomedicine operate on inputs that lack intrinsic order. Models may make use of gene expression data matrices from RNA sequencing or microarray experiments in which rows represent samples and columns represent genes. To account for relationships between genes, one might incorporate known interactions or correlations when making predictions or generating a low-dimensional representation of the data (Figure 2). This is comparable to the manner in which sequence context pushes models to consider nearby base pairs similarly.

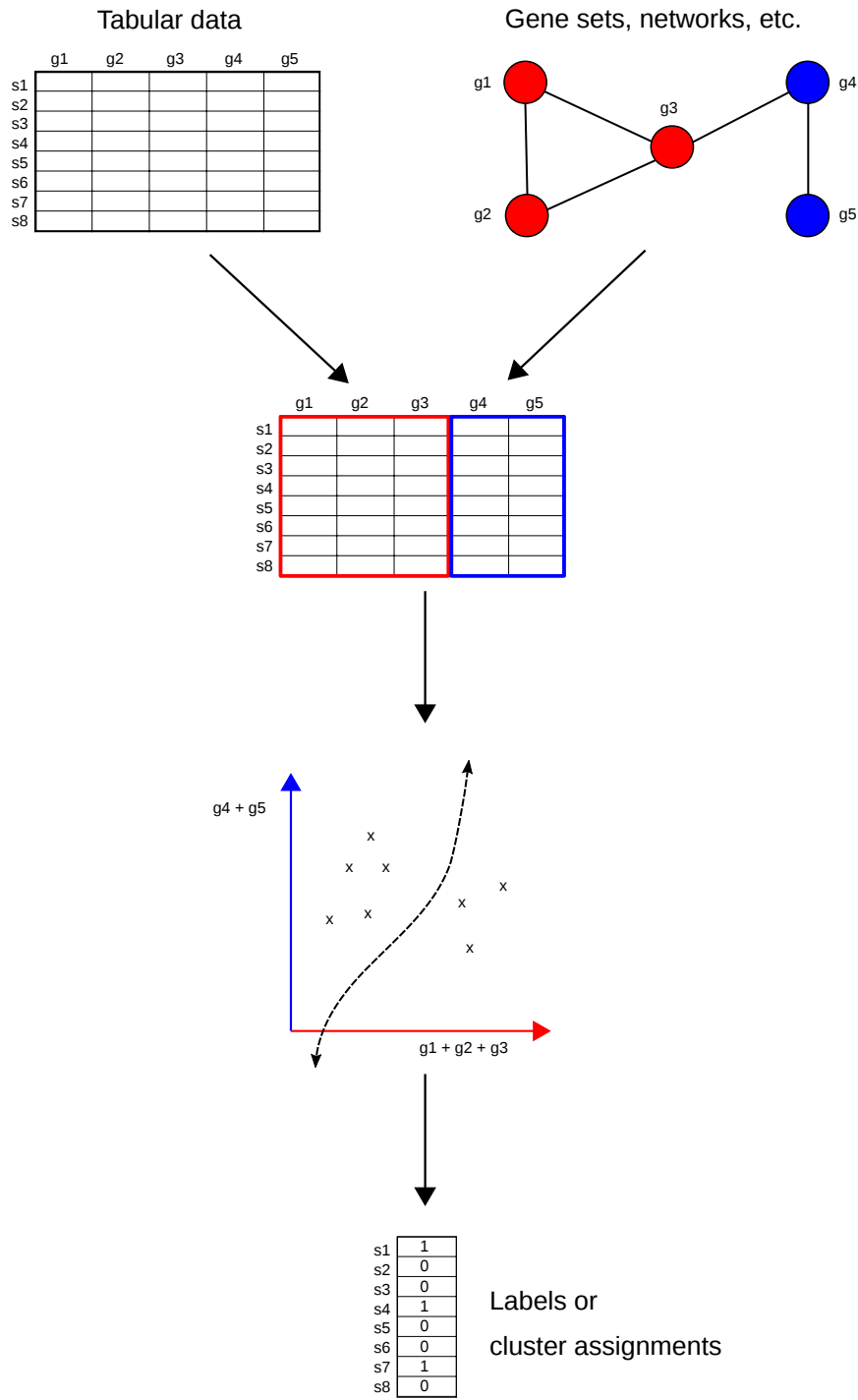

**Figure 2:** The relationships between genes provide structure that can be incorporated into machine learning models. One common approach is to use a network or collection of gene sets to embed the data in a lower-dimensional space, in which genes that are in the same gene sets or that are well-connected in the network have a similar representation in the lower-dimensional space. The embedded data can then be used for classification or clustering tasks.

## Applications in transcriptomics

Models built from gene expression data can benefit from incorporating gene-level relationships. One form that this knowledge commonly takes is a database of gene sets, which may represent biological pathways or gene signatures for a biological state of interest. PLIER [29] uses gene set information from MSigDB [30] and cell type markers to extract a representation of gene expression data that corresponds to biological processes and reduces technical noise. The resulting gene set-aligned representation accurately decomposed cell type mixtures. MultiPLIER [31] applied PLIER to the

recount2 gene expression compendium [32] to develop a model that shares information across multiple tissues and diseases, including rare diseases with limited sample sizes. PASNet [33] uses MSigDB to inform the structure of a neural network for predicting patient outcomes in glioblastoma multiforme (GBM) from gene expression data. This approach aids interpretation, as pathway nodes in the network with high weights can be inferred to correspond to certain pathways in GBM outcome prediction.

Gene-level relationships can also be represented with networks. Network nodes typically represent genes and real-valued edges may represent interactions or correlations between genes, often in a tissue or cell type context of interest. netNMF-sc [34] incorporates coexpression networks [35] as a smoothing term for dimension reduction and dropout imputation in single-cell gene expression data. The coexpression network improves performance for identifying cell types and of cell cycle marker genes, as compared to using raw gene expression or other single-cell dimension reduction methods. Combining gene expression data with a network-derived smoothing term also improved prediction of patient drug response in acute myeloid leukemia [36] and detection of mutated cancer genes [37]. PIMKL [38] combines network and pathway data to predict disease-free survival from breast cancer cohorts. This method takes as input both RNA-seq gene expression data and copy number alteration data, but can also be applied to gene expression data alone.

Gene regulatory networks can also augment models for gene expression data. These networks describe how the expression of genes is modulated by biological regulators such as transcription factors, microRNAs, or small molecules. creNET [39] integrates a gene regulatory network, derived from STRING [40], with a sparse logistic regression model to predict phenotypic response in clinical trials for ulcerative colitis and acute kidney rejection. The gene regulatory information allows the model to identify the biological regulators associated with the response, potentially giving mechanistic insight into differential clinical trial response. GRRANN [41], which was applied to the same data as creNET, uses a gene regulatory network to inform the structure of a neural network. Several other methods [42,43] have also used gene regulatory network structure to constrain the structure of a neural network, reducing the number of parameters to be fit and facilitating interpretation.

## Applications in genetics

Approaches that incorporate gene set or network structure into genetic studies have a long history [44,45]. Recent applications include expression quantitative trait loci (eQTL) mapping studies, which aim to identify associations between genetic variants and gene expression. netReg [46] implements a graph-regularized dual LASSO algorithm for eQTL mapping [47] in a publicly available R package. This model smooths regression coefficients simultaneously based on networks describing associations between genes (target variables in the eQTL regression model) and between variants (predictors in the eQTL regression model). eQTL information is also used in conjunction with genetic variant information to predict phenotypes, in an approach known as Mendelian randomization (MR). In [48], a smoothing term derived from a gene regulatory network is used in an MR model. The model with the network smoothing term, applied to a human liver dataset, more robustly identifies genes that influence enzyme activity than a network-agnostic model. As genetic datasets grow, we expect that researchers will continue to develop models that leverage gene set and network databases.

## Other models incorporating biological structure

Knowledge about biological entities is often organized in an ontology, which is a directed graph that encodes relationships between entities (see Figure 3 for a visual example). The Gene Ontology (GO) [3] describes the relationships between cellular subsystems and other attributes describing proteins or genes. DCell [49] uses GO to inform the connectivity of a neural network predicting the effects of gene deletions on yeast growth. DCell performs comparably to an unconstrained neural network for this task. Additionally, it is easier to interpret: a cellular subsystem with high neuron outputs under a particular gene deletion can be inferred to be strongly affected by the gene deletion, providing a putative genotype-phenotype association. DeepGO [50] uses a similar approach to predict protein function from amino acid sequence with a neural network constrained by the dependencies of GO. However, a follow-up paper by the same authors [51] showed that this hierarchy-aware approach can be outperformed by a hierarchy-naive CNN, which uses only amino acid sequence and similarity to labeled training set proteins. This suggests a tradeoff between interpretability and predictive accuracy for protein function prediction.

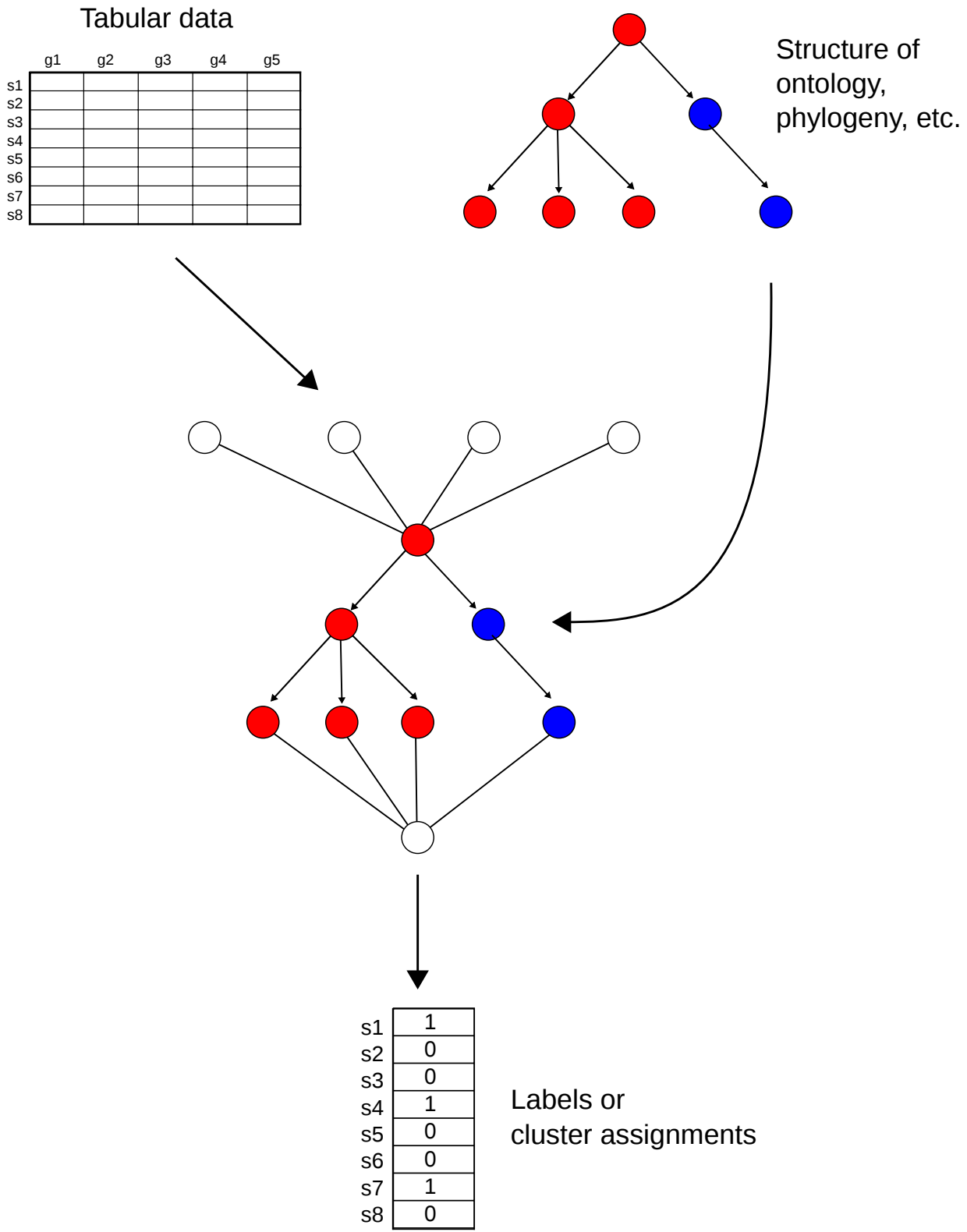

**Figure 3:** Directed graph-structured data, such as an ontology or phylogenetic tree can be incorporated into machine learning models. Here, the connections in the neural network used to predict a set of labels parallel those in the tree graph. This type of constraint can also be useful in model interpretation: for example, if the red-shaded nodes have high neuron outputs for a given sample, then the subsystem encoded in the red-shaded part of the tree graph is most likely important in making predictions for that sample.

Phylogenetic trees, or hierarchies describing the evolutionary relationships between species, can be useful for a similar purpose. glmmTree [52] uses a phylogenetic tree describing the relationship between microorganisms to improve predictions of age based on gut microbiome data. The same authors combine a similar phylogeny smoothing strategy with sparse regression to model caffeine intake and smoking status based on microbiome data [53]. Phylogenetic trees can also describe the relationships between subclones of a tumor, which are fundamental to understanding cancer evolution and development. Using a tumor phylogeny inferred from copy number aberration (CNA) sequencing data as a smoothing term for deconvolving tumor subclones provided more robust predictions than a phylogeny-free model [54]. The tree structure of the phylogeny and the subclone mixture model are fit jointly to the CNA data.

Depending on the application, other forms of structure or prior knowledge can inform predictions and interpretation of the model's output. CYCLOPS [55] uses a circular node autoencoder [56] to order periodic gene expression data and estimate circadian rhythms. The authors validated the method by correctly ordering samples without temporal labels and identifying genes with known circadian expression. They then applied it to compare gene expression in normal and cancerous liver biopsies, identifying drug targets with circadian expression as candidates for chronotherapy. NetBiTE [57] uses drug-gene interaction information from GDSC [58], in addition to protein interaction data, to build a tree ensemble model with splits that are biased toward high-confidence drug-gene interactions. The model predicts sensitivity to drugs that inhibit critical signaling pathways in cancer, showing improved predictive performance compared to random forests, another commonly used tree ensemble model.

## Conclusions and future directions

As the quantity and richness of biomedical data has increased, sequence repositories and interaction databases have expanded and become more robust. This raises opportunities to integrate these resources into the structure of machine learning models. Going forward, there is an outstanding need for benchmarks comparing these approaches across diverse datasets and prediction problems, along the lines of the evaluation in [59] but updated and expanded to include recent methods and applications. Improved benchmarking should lead to a better understanding of which dataset characteristics align with which approaches.

Many methods described in this review have open-source implementations available; however, increased availability of performant and extensible implementations of these models and algorithms would facilitate further use and development. In the future, we foresee that incorporating structured biomedical data will become commonplace for improving model interpretability and boosting performance when sample size is limited.

## Acknowledgements

The authors would like to thank Daniel Himmelstein for a critical reading of the manuscript and helpful discussion.

## Reference Annotations

[**] Annotation for BPNet [16]:

This paper describes BPNet, a neural network for predicting transcription factor (TF) binding profiles from raw DNA sequence. The model is able to accurately infer the spacing and periodicity of pluripotency-related TFs in mouse embryonic stem cells, leading to an improved understanding of the motif syntax of combinatorial TF binding in cell development.

[*] Annotation for cDeepbind [18]:

cDeepbind is a neural network model for predicting RNA binding protein (RBP) specificity from RNA sequence and secondary structure information. The authors show that this combined approach provides an improvement over previous models that use only sequence information.

[*] Annotation for DeepDiff [23]:

DeepDiff uses a long short-term memory neural network to predict differential gene expression from the spatial structure of histone modification measurements. The network has a multi-task objective, enabling gene expression predictions to be made simultaneously in multiple cell types.

[**] Annotation for DeepVariant [24]:

This paper describes DeepVariant, a neural network model for distinguishing true genetic variants from errors in next-generation DNA sequencing data. The model adapts techniques from the image processing community to fit a model on images of read pileups around candidate variants, using information about the sequence around the candidate variant site to make predictions about the true genotype at the site.

[**] Annotation for PLIER [29]:

This paper describes a "pathway-level information extractor" (PLIER), a method for reducing the dimension of gene expression data in a manner that aligns with known biological pathways or informative gene sets. The method can also reduce the effects of technical noise. The authors show that PLIER can be used to improve cell type inference and as a component in eQTL studies.

[**] Annotation for netNMF-sc [34]:

netNMF-sc is a dimension reduction method that uses network information to "smooth" a matrix factorization of single-cell gene expression data, such that genes that are connected in the network have a similar low-dimensional representation. Inclusion of network information is particularly useful when analyzing single-cell expression data, due to its ability to mitigate "dropouts" and other sources of variability that are present at the single cell level.

[*] Annotation for Attribution Priors [36]:

This paper describes "model attribution priors", or a method for constraining a machine learning model's behavior during training with prior beliefs or expectations about the data or problem structure. As an example of this concept, the authors show that incorporation of network data improves the performance of a model for drug response prediction in acute myeloid leukemia.

[*] Annotation for PIMKL [38]:

In this paper, the authors present an algorithm for combining gene expression and copy number data with prior information, such as gene networks and pathways or gene set annotations, to predict survival in breast cancer. The weights learned by the model are also interpretable, providing a putative set of explanatory features for the prediction task.

[**] Annotation for creNET [39]:

This work describes creNET, a regression model for gene expression data that uses information about gene regulation to differentially weight or penalize gene sets that are co-regulated. The authors show that the model can be used to predict phenotype from gene expression data in clinical trials. The model also provides interpretable weights for each gene regulator.

[**] Annotation for DCell [49]:

This paper presents DCell, a neural network model for prediction of yeast growth phenotype from gene deletions. The structure of the neural network is constrained by the relationships encoded in the Gene Ontology (GO), enabling predictions for a given input to be interpreted based on the subsystems of GO that they activate. Thus, the neural network can be seen as connecting genotype to phenotype.

[*] Annotation for DeepGO [50]:

Here, the authors describe a method for predicting protein function from amino acid sequence, incorporating the dependency structure of the Gene Ontology (GO) into their neural network used for prediction. Using the GO information provides a performance improvement over similar models that do not incorporate this information.

[**] Annotation for NetBITE [57]:

This paper describes a method for using prior knowledge about drug targets to inform the structure of a tree ensemble model, used for predicting IC50 drug sensitivity from gene expression data. The model also uses a protein interaction network to "smooth" the gene weights, such that genes that are related in the network will have a similar influence on predictions.

# References

1. **Network propagation: a universal amplifier of genetic associations**
Lenore Cowen, Trey Ideker, Benjamin J. Raphael, Roded Sharan
*Nature Reviews Genetics* (2017-06-12) <https://doi.org/gbhkwn>
DOI: [10.1038/nrg.2017.38](https://doi.org/10.1038/nrg.2017.38) · PMID: [28607512](https://www.ncbi.nlm.nih.gov/pubmed/28607512)

2. **Visible Machine Learning for Biomedicine**
Michael K. Yu, Jianzhu Ma, Jasmin Fisher, Jason F. Kreisberg, Benjamin J. Raphael, Trey Ideker
*Cell* (2018-06) <https://doi.org/gdqcd8>
DOI: [10.1016/j.cell.2018.05.056](https://doi.org/10.1016/j.cell.2018.05.056) · PMID: [29906441](https://www.ncbi.nlm.nih.gov/pubmed/29906441) · PMCID: [PMC6483071](https://www.ncbi.nlm.nih.gov/pmc/articles/PMC6483071)

3. **The Gene Ontology Resource: 20 years and still GOing strong***Nucleic Acids Research* (2018-11-05) <https://doi.org/gf63mb>
DOI: [10.1093/nar/gky1055](https://doi.org/10.1093/nar/gky1055) · PMID: [30395331](https://www.ncbi.nlm.nih.gov/pubmed/30395331) · PMCID: [PMC6323945](https://www.ncbi.nlm.nih.gov/pmc/articles/PMC6323945)

4. **Machine Learning in Genomic Medicine: A Review of Computational Problems and Data Sets**
Michael K. K. Leung, Andrew Delong, Babak Alipanahi, Brendan J. Frey
*Proceedings of the IEEE* (2016-01) <https://doi.org/f75grb>
DOI: [10.1109/jproc.2015.2494198](https://doi.org/10.1109/jproc.2015.2494198)

5. **Diet Networks: Thin Parameters for Fat Genomics**
Adriana Romero, Pierre Luc Carrier, Akram Erraqabi, Tristan Sylvain, Alex Auvolat, Etienne Dejoie, Marc-André Legault, Marie-Pierre Dubé, Julie G. Hussin, Yoshua Bengio
*arXiv* (2016-11-28) <https://arxiv.org/abs/1611.09340v3>

6. **ImageNet: A large-scale hierarchical image database**
Jia Deng, Wei Dong, Richard Socher, Li-Jia Li, Kai Li, Li Fei-Fei
*2009 IEEE Conference on Computer Vision and Pattern Recognition* (2009-06) <https://doi.org/cvc7xp>
DOI: [10.1109/cvpr.2009.5206848](https://doi.org/10.1109/cvpr.2009.5206848)

7. **Learning Word Vectors for Sentiment Analysis**
Andrew Maas, Raymond E. Daly, Peter T. Pham, Dan Huang, Andrew Y. Ng, Christopher Potts
(2011-06) <https://www.aclweb.org/anthology/P11-1015>

8. **To Embed or Not: Network Embedding as a Paradigm in Computational Biology**
Walter Nelson, Marinka Zitnik, Bo Wang, Jure Leskovec, Anna Goldenberg, Roded Sharan
*Frontiers in Genetics* (2019-05-01) <https://doi.org/gf8rdf>
DOI: [10.3389/fgene.2019.00381](https://doi.org/10.3389/fgene.2019.00381) · PMID: [31118945](https://www.ncbi.nlm.nih.gov/pubmed/31118945) · PMCID: [PMC6504708](https://www.ncbi.nlm.nih.gov/pmc/articles/PMC6504708)

9. **Machine learning for integrating data in biology and medicine: Principles, practice, and opportunities**
Marinka Zitnik, Francis Nguyen, Bo Wang, Jure Leskovec, Anna Goldenberg, Michael M. Hoffman
*Information Fusion* (2019-10) <https://doi.org/gf7rj8>
DOI: [10.1016/j.inffus.2018.09.012](https://doi.org/10.1016/j.inffus.2018.09.012) · PMID: [30467459](https://www.ncbi.nlm.nih.gov/pubmed/30467459) · PMCID: [PMC6242341](https://www.ncbi.nlm.nih.gov/pmc/articles/PMC6242341)

10. **DEEP: a general computational framework for predicting enhancers**
Dimitrios Kleftogiannis, Panos Kalnis, Vladimir B. Bajic
*Nucleic Acids Research* (2014-11-05) <https://doi.org/gcgk83>
DOI: [10.1093/nar/gku1058](https://doi.org/10.1093/nar/gku1058) · PMID: [25378307](https://www.ncbi.nlm.nih.gov/pubmed/25378307) · PMCID: [PMC4288148](https://www.ncbi.nlm.nih.gov/pmc/articles/PMC4288148)

11. **The human splicing code reveals new insights into the genetic determinants of disease**
H. Y. Xiong, B. Alipanahi, L. J. Lee, H. Bretschneider, D. Merico, R. K. C. Yuen, Y. Hua, S. Gueroussov, H. S. Najafabadi, T. R. Hughes, … B. J. Frey
*Science* (2014-12-18) https://doi.org/f6wzj2
DOI: 10.1126/science.1254806 · PMID: 25525159 · PMCID: PMC4362528

12. **Deep learning for computational biology**
Christof Angermueller, Tanel Pärnamaa, Leopold Parts, Oliver Stegle
*Molecular Systems Biology* (2016-07) https://doi.org/f8xtvh
DOI: 10.15252/msb.20156651 · PMID: 27474269 · PMCID: PMC4965871

13. **Predicting the sequence specificities of DNA- and RNA-binding proteins by deep learning**
Babak Alipanahi, Andrew Delong, Matthew T Weirauch, Brendan J Frey
*Nature Biotechnology* (2015-07-27) https://doi.org/f7mkrd
DOI: 10.1038/nbt.3300 · PMID: 26213851

14. **Predicting effects of noncoding variants with deep learning–based sequence model**
Jian Zhou, Olga G Troyanskaya
*Nature Methods* (2015-08-24) https://doi.org/gcgk8g
DOI: 10.1038/nmeth.3547 · PMID: 26301843 · PMCID: PMC4768299

15. **Basset: learning the regulatory code of the accessible genome with deep convolutional neural networks**
David R. Kelley, Jasper Snoek, John L. Rinn
*Genome Research* (2016-05-03) https://doi.org/f8sw35
DOI: 10.1101/gr.200535.115 · PMID: 27197224 · PMCID: PMC4937568

16. **Deep learning at base-resolution reveals motif syntax of the cis-regulatory code**
Žiga Avsec, Melanie Weilert, Avanti Shrikumar, Amr Alexandari, Sabrina Krueger, Khyati Dalal, Robin Fropf, Charles McAnany, Julien Gagneur, Anshul Kundaje, Julia Zeitlinger
*Cold Spring Harbor Laboratory* (2019-08-21) https://doi.org/gf64fc
DOI: 10.1101/737981

17. **Multi-Scale Context Aggregation by Dilated Convolutions**
Fisher Yu, Vladlen Koltun
*arXiv* (2015-11-23) https://arxiv.org/abs/1511.07122v3

18. **cDeepbind: A context sensitive deep learning model of RNA-protein binding**
Shreshth Gandhi, Leo J. Lee, Andrew Delong, David Duvenaud, Brendan J. Frey
*Cold Spring Harbor Laboratory* (2018-06-12) https://doi.org/gf68nk
DOI: 10.1101/345140

19. **A Deep Neural Network for Predicting and Engineering Alternative Polyadenylation**
Nicholas Bogard, Johannes Linder, Alexander B. Rosenberg, Georg Seelig
*Cell* (2019-06) https://doi.org/gf66nc
DOI: 10.1016/j.cell.2019.04.046 · PMID: 31178116 · PMCID: PMC6599575

20. **DeepSignal: detecting DNA methylation state from Nanopore sequencing reads using deep-learning**
Peng Ni, Neng Huang, Zhi Zhang, De-Peng Wang, Fan Liang, Yu Miao, Chuan-Le Xiao, Feng Luo, Jianxin Wang
*Bioinformatics* (2019-04-17) https://doi.org/gf66qw
DOI: 10.1093/bioinformatics/btz276 · PMID: 30994904

21. **MRCNN: a deep learning model for regression of genome-wide DNA methylation**
Qi Tian, Jianxiao Zou, Jianxiong Tang, Yuan Fang, Zhongli Yu, Shicai Fan
*BMC Genomics* (2019-04) https://doi.org/gf48g6
DOI: 10.1186/s12864-019-5488-5 · PMID: 30967120 · PMCID: PMC6457069

22. **Attend and Predict: Understanding Gene Regulation by Selective Attention on Chromatin**
Ritambhara Singh, Jack Lanchantin, Arshdeep Sekhon, Yanjun Qi
*Cold Spring Harbor Laboratory* (2018-05-25) https://doi.org/gf66qz
DOI: 10.1101/329334

23. **DeepDiff: DEEP-learning for predicting DIFFerential gene expression from histone modifications**
Arshdeep Sekhon, Ritambhara Singh, Yanjun Qi
*Bioinformatics* (2018-09-01) https://doi.org/gd9mk4
DOI: 10.1093/bioinformatics/bty612 · PMID: 30423076

24. **A universal SNP and small-indel variant caller using deep neural networks**
Ryan Poplin, Pi-Chuan Chang, David Alexander, Scott Schwartz, Thomas Colthurst, Alexander Ku, Dan Newburger, Jojo Dijamco, Nam Nguyen, Pegah T Afshar, … Mark A DePristo
*Nature Biotechnology* (2018-09-24) https://doi.org/gd8gkf
DOI: 10.1038/nbt.4235 · PMID: 30247488

25. **A multi-task convolutional deep neural network for variant calling in single molecule sequencing**
Ruibang Luo, Fritz J. Sedlazeck, Tak-Wah Lam, Michael C. Schatz
*Nature Communications* (2019-03-01) https://doi.org/gf4c37
DOI: 10.1038/s41467-019-09025-z · PMID: 30824707 · PMCID: PMC6397153

26. **SICaRiO: Short Indel Call filteRing with bOosting**
Md Shariful Islam Bhuyan, Itsik Pe'er, M. Sohel Rahman
*Cold Spring Harbor Laboratory* (2019-04-07) https://doi.org/gf68d6
DOI: 10.1101/601450

27. **Machine learning-based detection of insertions and deletions in the human genome**
Charles Curnin, Rachel L. Goldfeder, Shruti Marwaha, Devon Bonner, Daryl Waggott, Matthew T. Wheeler, Euan A. Ashley,
*Cold Spring Harbor Laboratory* (2019-05-05) https://doi.org/gf68d7
DOI: 10.1101/628222

28. **Deep convolutional neural networks for accurate somatic mutation detection**
Sayed Mohammad Ebrahim Sahraeian, Ruolin Liu, Bayo Lau, Karl Podesta, Marghoob Mohiyuddin, Hugo Y. K. Lam
*Nature Communications* (2019-03-04) https://doi.org/gf68f8
DOI: 10.1038/s41467-019-09027-x · PMID: 30833567 · PMCID: PMC6399298

29. **Pathway-level information extractor (PLIER) for gene expression data**
Weiguang Mao, Elena Zaslavsky, Boris M. Hartmann, Stuart C. Sealfon, Maria Chikina
*Nature Methods* (2019-06-27) https://doi.org/gf75g6
DOI: 10.1038/s41592-019-0456-1 · PMID: 31249421

30. **Gene set enrichment analysis: A knowledge-based approach for interpreting genome-wide expression profiles**
A. Subramanian, P. Tamayo, V. K. Mootha, S. Mukherjee, B. L. Ebert, M. A. Gillette, A. Paulovich, S. L.

Pomeroy, T. R. Golub, E. S. Lander, J. P. Mesirov
*Proceedings of the National Academy of Sciences* (2005-09-30) https://doi.org/d4qbh8
DOI: 10.1073/pnas.0506580102 · PMID: 16199517 · PMCID: PMC1239896

31. **MultiPLIER: A Transfer Learning Framework for Transcriptomics Reveals Systemic Features of Rare Disease**
Jaclyn N. Taroni, Peter C. Grayson, Qiwen Hu, Sean Eddy, Matthias Kretzler, Peter A. Merkel, Casey S. Greene
*Cell Systems* (2019-05) https://doi.org/gf75g5
DOI: 10.1016/j.cels.2019.04.003 · PMID: 31121115 · PMCID: PMC6538307

32. **Reproducible RNA-seq analysis using recount2**
Leonardo Collado-Torres, Abhinav Nellore, Kai Kammers, Shannon E Ellis, Margaret A Taub, Kasper D Hansen, Andrew E Jaffe, Ben Langmead, Jeffrey T Leek
*Nature Biotechnology* (2017-04) https://doi.org/gf75hp
DOI: 10.1038/nbt.3838 · PMID: 28398307 · PMCID: PMC6742427

33. **PASNet: pathway-associated sparse deep neural network for prognosis prediction from high-throughput data**
Jie Hao, Youngsoon Kim, Tae-Kyung Kim, Mingon Kang
*BMC Bioinformatics* (2018-12) https://doi.org/gf75g9
DOI: 10.1186/s12859-018-2500-z · PMID: 30558539 · PMCID: PMC6296065

34. **netNMF-sc: Leveraging gene-gene interactions for imputation and dimensionality reduction in single-cell expression analysis**
Rebecca Elyanow, Bianca Dumitrascu, Barbara E. Engelhardt, Benjamin J. Raphael
*Cold Spring Harbor Laboratory* (2019-02-08) https://doi.org/gf386x
DOI: 10.1101/544346

35. **COEXPEDIA: exploring biomedical hypotheses via co-expressions associated with medical subject headings (MeSH)**
Sunmo Yang, Chan Yeong Kim, Sohyun Hwang, Eiru Kim, Hyojin Kim, Hongseok Shim, Insuk Lee
*Nucleic Acids Research* (2016-09-26) https://doi.org/f9v38k
DOI: 10.1093/nar/gkw868 · PMID: 27679477 · PMCID: PMC5210615

36. **Learning Explainable Models Using Attribution Priors**
Gabriel Erion, Joseph D. Janizek, Pascal Sturmfels, Scott Lundberg, Su-In Lee
*arXiv* (2019-06-25) https://arxiv.org/abs/1906.10670v1

37. **A novel network regularized matrix decomposition method to detect mutated cancer genes in tumour samples with inter-patient heterogeneity**
Jianing Xi, Ao Li, Minghui Wang
*Scientific Reports* (2017-06-06) https://doi.org/gcq9j7
DOI: 10.1038/s41598-017-03141-w · PMID: 28588243 · PMCID: PMC5460199

38. **PIMKL: Pathway-Induced Multiple Kernel Learning**
Matteo Manica, Joris Cadow, Roland Mathis, María Rodríguez Martínez
*npj Systems Biology and Applications* (2019-03-05) https://doi.org/gf8ck6
DOI: 10.1038/s41540-019-0086-3 · PMID: 30854223 · PMCID: PMC6401099

39. **Robust phenotype prediction from gene expression data using differential shrinkage of co-regulated genes**
Kourosh Zarringhalam, David Degras, Christoph Brockel, Daniel Ziemek

*Scientific Reports* (2018-01-19) https://doi.org/gcwzdn
DOI: 10.1038/s41598-018-19635-0 · PMID: 29352257 · PMCID: PMC5775343

40. **STRING v10: protein–protein interaction networks, integrated over the tree of life**
Damian Szklarczyk, Andrea Franceschini, Stefan Wyder, Kristoffer Forslund, Davide Heller, Jaime Huerta-Cepas, Milan Simonovic, Alexander Roth, Alberto Santos, Kalliopi P. Tsafou, … Christian von Mering
*Nucleic Acids Research* (2014-10-28) https://doi.org/f64rfn
DOI: 10.1093/nar/gku1003 · PMID: 25352553 · PMCID: PMC4383874

41. **A biological network-based regularized artificial neural network model for robust phenotype prediction from gene expression data**
Tianyu Kang, Wei Ding, Luoyan Zhang, Daniel Ziemek, Kourosh Zarringhalam
*BMC Bioinformatics* (2017-12) https://doi.org/gf8cm6
DOI: 10.1186/s12859-017-1984-2 · PMID: 29258445 · PMCID: PMC5735940

42. **Using neural networks for reducing the dimensions of single-cell RNA-Seq data**
Chieh Lin, Siddhartha Jain, Hannah Kim, Ziv Bar-Joseph
*Nucleic Acids Research* (2017-07-31) https://doi.org/gcnzb7
DOI: 10.1093/nar/gkx681 · PMID: 28973464 · PMCID: PMC5737331

43. **Genetic Neural Networks: an artificial neural network architecture for capturing gene expression relationships**
Ameen Eetemadi, Ilias Tagkopoulos
*Bioinformatics* (2018-11-19) https://doi.org/gfnks6
DOI: 10.1093/bioinformatics/bty945 · PMID: 30452523

44. **Nonparametric pathway-based regression models for analysis of genomic data**
Z. Wei, H. Li
*Biostatistics* (2006-06-13) https://doi.org/fdmgqm
DOI: 10.1093/biostatistics/kxl007 · PMID: 16772399

45. **Network-constrained regularization and variable selection for analysis of genomic data**
C. Li, H. Li
*Bioinformatics* (2008-03-01) https://doi.org/fk8n4b
DOI: 10.1093/bioinformatics/btn081 · PMID: 18310618

46. **netReg: network-regularized linear models for biological association studies**
Simon Dirmeier, Christiane Fuchs, Nikola S Mueller, Fabian J Theis
*Bioinformatics* (2017-10-25) https://doi.org/gcg9xq
DOI: 10.1093/bioinformatics/btx677 · PMID: 29077797 · PMCID: PMC6030897

47. **Graph-regularized dual Lasso for robust eQTL mapping**
Wei Cheng, Xiang Zhang, Zhishan Guo, Yu Shi, Wei Wang
*Bioinformatics* (2014-06-11) https://doi.org/f58j6m
DOI: 10.1093/bioinformatics/btu293 · PMID: 24931977 · PMCID: PMC4058913

48. **Integrative analysis of genetical genomics data incorporating network structures**
Bin Gao, Xu Liu, Hongzhe Li, Yuehua Cui
*Biometrics* (2019-04-29) https://doi.org/gf8f9q
DOI: 10.1111/biom.13072 · PMID: 31009063

49. **Using deep learning to model the hierarchical structure and function of a cell**
Jianzhu Ma, Michael Ku Yu, Samson Fong, Keiichiro Ono, Eric Sage, Barry Demchak, Roded Sharan, Trey Ideker
*Nature Methods* (2018-03-05) https://doi.org/gc46jp
DOI: 10.1038/nmeth.4627 · PMID: 29505029 · PMCID: PMC5882547

50. **DeepGO: predicting protein functions from sequence and interactions using a deep ontology-aware classifier**
Maxat Kulmanov, Mohammed Asif Khan, Robert Hoehndorf
*Bioinformatics* (2017-10-03) https://doi.org/gc3nb8
DOI: 10.1093/bioinformatics/btx624 · PMID: 29028931 · PMCID: PMC5860606

51. **DeepGOPlus: improved protein function prediction from sequence**
Maxat Kulmanov, Robert Hoehndorf
*Bioinformatics* (2019-07-27) https://doi.org/gf84d8
DOI: 10.1093/bioinformatics/btz595 · PMID: 31350877

52. **Predictive Modeling of Microbiome Data Using a Phylogeny-Regularized Generalized Linear Mixed Model**
Jian Xiao, Li Chen, Stephen Johnson, Yue Yu, Xianyang Zhang, Jun Chen
*Frontiers in Microbiology* (2018-06-27) https://doi.org/gdtz4z
DOI: 10.3389/fmicb.2018.01391 · PMID: 29997602 · PMCID: PMC6030386

53. **A Phylogeny-Regularized Sparse Regression Model for Predictive Modeling of Microbial Community Data**
Jian Xiao, Li Chen, Yue Yu, Xianyang Zhang, Jun Chen
*Frontiers in Microbiology* (2018-12-19) https://doi.org/gf8qcc
DOI: 10.3389/fmicb.2018.03112 · PMID: 30619188 · PMCID: PMC6305753

54. **Tumor Copy Number Deconvolution Integrating Bulk and Single-Cell Sequencing Data**
Haoyun Lei, Bochuan Lyu, E. Michael Gertz, Alejandro A. Schäffer, Xulian Shi, Kui Wu, Guibo Li, Liqin Xu, Yong Hou, Michael Dean, Russell Schwartz
*Lecture Notes in Computer Science* (2019) https://doi.org/gf8qck
DOI: 10.1007/978-3-030-17083-7_11

55. **CYCLOPS reveals human transcriptional rhythms in health and disease**
Ron C. Anafi, Lauren J. Francey, John B. Hogenesch, Junhyong Kim
*Proceedings of the National Academy of Sciences* (2017-04-24) https://doi.org/f9796k
DOI: 10.1073/pnas.1619320114 · PMID: 28439010 · PMCID: PMC5441789

56. **Circular Nodes in Neural Networks**
Michael J. Kirby, Rick Miranda
*Neural Computation* (1996-02-15) https://doi.org/ffcww8
DOI: 10.1162/neco.1996.8.2.390

57. **Network-based Biased Tree Ensembles (NetBiTE) for Drug Sensitivity Prediction and Drug Sensitivity Biomarker Identification in Cancer**
Ali Oskooei, Matteo Manica, Roland Mathis, Maria Rodriguez Martinez
*arXiv* (2018-08-18) https://arxiv.org/abs/1808.06603v2

58. **Genomics of Drug Sensitivity in Cancer (GDSC): a resource for therapeutic biomarker discovery in cancer cells**
Wanjuan Yang, Jorge Soares, Patricia Greninger, Elena J. Edelman, Howard Lightfoot, Simon Forbes,

Nidhi Bindal, Dave Beare, James A. Smith, I. Richard Thompson, … Mathew J. Garnett
*Nucleic Acids Research* (2012-11-22) <https://doi.org/f4jrdn>
DOI: [10.1093/nar/gks1111](10.1093/nar/gks1111) · PMID: [23180760](23180760) · PMCID: [PMC3531057](PMC3531057)

59. **A Critical Evaluation of Network and Pathway-Based Classifiers for Outcome Prediction in Breast Cancer**
Christine Staiger, Sidney Cadot, Raul Kooter, Marcus Dittrich, Tobias Müller, Gunnar W. Klau, Lodewyk F. A. Wessels
*PLoS ONE* (2012-04-27) <https://doi.org/gf8rgw>
DOI: [10.1371/journal.pone.0034796](10.1371/journal.pone.0034796) · PMID: [22558100](22558100) · PMCID: [PMC3338754](PMC3338754)